\documentstyle[epsf]{elsart}
\begin{document}
\begin{frontmatter}
\title{\bf 
Local Thermodynamical Equilibrium and the Equation of State 
of Hot, Dense Matter Created in Au+Au Collisions at AGS }
\author{L.V. Bravina$^{a,1,2}$}, \author{M.I. Gorenstein$^{a,b,3}$}, 
\author{M. Belkacem$^{a,1}$}, \author{S.A. Bass$^{c,4}$}, 
\author{M. Bleicher$^a$}, \author{M. Brandstetter$^a$}, 
\author{M. Hofmann$^{a}$}, 
\author{S. Soff$^{a,d}$}, \author{C. Spieles$^{e,4}$}, \author{H. Weber$^{a}$},
\author{H. St{\"o}cker$^a$} and \author{W. Greiner$^a$ }
\address{
$^a$Institut f\"ur Theoretische Physik, Goethe Universit\"at
    Frankfurt, Germany\\
$^b$School of Physics and Astronomy, Tel Aviv University, 
    Tel Aviv, Israel\\
$^c$Dept. of Physics, Duke University, Durham,  USA\\
$^d$Gesellschaft f\"ur Schwerionenforschung, Darmstadt, Germany\\
$^e$Nuclear Science Division, LBNL, Berkeley, CA 94720, USA \\
$^1$Alexander von Humboldt Foundation Fellow \\
$^2$on leave of absence from \\
    the Institute for Nuclear Physics, Moscow State University, Russia \\
$^3$Permanent address: Bogolyubov Institute for Theoretical Physics, 
    Kiev, Ukraine\\
$^4$Feodor Lynen Fellow of the Alexander von Humboldt Foundation
}
\maketitle

\begin{abstract}
Local kinetic and chemical equilibration is studied 
for Au+Au collisions at 10.7 AGeV 
in the microscopic 
Ultrarelativistic
Quantum Molecular Dynamics model (UrQMD).
The UrQMD model exhibits dramatic deviations from equilibrium 
during the high density  phase of the collision.
Thermal and chemical equilibration 
of the hadronic matter seems to be established in the later stages
during a quasi-isentropic expansion, observed 
in the central reaction cell with volume 125 fm$^{3}$. 
For $t\geq$10~fm/c the hadron energy spectra in the cell
are nicely reproduced by Boltzmann distributions with a common
rapidly dropping temperature. Hadron yields change drastically and 
at the late expansion stage follow closely those of an ideal gas 
statistical model. 
The equation of state seems to be simple at late times:
$P\cong0.12\varepsilon$. The time evolution of other thermodynamical 
variables in the cell is  also presented. 
\end{abstract}

\end{frontmatter}

\newpage
\section{Introduction}

The main  goal of the relativistic heavy-ion experiments at Brookhaven and
CERN is to study the properties, e.g. 
the equation of state (EoS), of strongly interacting 
hot and dense hadronic matter produced in the course
of nuclear collisions. 
A question of great importance in these studies is to determine the EoS and 
whether local (LTE) or even global thermodynamical equilibrium 
is reached in the system or not. 
This is a crucial point for
a large group of models, widely used for data  analysis.
Despite the long history of such investigations 
(e.g. \cite{Fermi,Pomeranchuk,Landau,Shuryak,Hofmann,Geiger,Heinz}), 
this  question remains still open.

One can try to check  the assumption
of LTE 
by analyzing the experimental particle number ratios and hadron
momentum spectra 
(e.g. \cite{cleym,braun,goren,flow}). 
Due to the time duration of chemical (inelastic) and 
kinetic (elastic) freeze-out of the 
system in A+A collisions 
\cite{Bass,bravina95,sorge95,Matiello9597}, 
the answer is not conclusive.
In contrast to the microscopic approaches \cite{urqmd1,QGSM,RQMD,ARC,ART},
macroscopic models like hydrodynamical \cite{Twofl,BMGR,Strott} 
or thermal \cite{Heinz,braun,goren} models adopt 
LTE as ad hoc assumption, use the Equation of State \cite{Greiner} as
a parametrization and apply usually an instantaneous freeze-out 
of hadrons, produced from many different space-time cells.
The freeze-out conditions for these cells -- temperatures, baryonic
chemical potentials and collective velocities -- are actually quite different
and model dependent (see e.g. \cite{flow,Matiello9597,Bass1}). 
Also note that the experimental 
particle spectra in elementary $e^+e^-$ and hadron-hadron
collisions at high energies already look "thermalized" \cite{Bass1,Becattini}.

Because of these difficulties 
it is very promising to check the LTE and to  extract the EoS 
at different stages of relativistic heavy ion reactions 
based on a detailed microscopic
analysis. 
Here, we employ  for a such analysis 
the microscopic Ultrarelativistic
Quantum Molecular Dynamics model (UrQMD) \cite{urqmd1}. 

\section{Ultrarelativistic Quantum Molecular Dynamics}

UrQMD  is a N-body transport model designed 
for describing  heavy ion collisions
in the laboratory energy range from several MeV
to several TeV per nucleon. 
The detailed presentation of UrQMD and underlying concepts together with a 
comparison with experimental data
is now publically accessible \cite{urqmd1}. 
The model treats binary 
elastic and inelastic hadronic collisions and many-body resonance
decays.
The inelastic collisions and decays are the only source for changing  the
chemical  composition of the system, 
while elastic collisions change only the momentum distributions of the hadrons.
Both processes drive the system  towards thermodynamical  equilibrium. 
However, the time scales may be too short in heavy ion collisions 
for actually achieving LTE.

The UrQMD model is based on string and resonance excitation both 
in primary nucleon-nucleon collisions from projectile and target, 
some also in secondary hadronic collisions. 
There are  55 baryon and 32 meson states  as discrete degrees 
of freedom in the model 
as well as their anti-particles and explicit isospin-projected states 
with masses up to 2.5 GeV. 
From 1.5 GeV the strings can be populated as continuous
degrees of freedom. 
The experimental hadron cross sections and resonance 
decay widths are used when possible.
At high energies a string mechanism is implemented to 
simulate soft hadronic interactions.  
Hadrons, produced
through the string decays, have non-zero formation
time which depends on the four-momentum of the particle. 
Newly produced particles cannot interact during their formation time.
Leading hadrons (containing the constituent quarks) 
interact within their 
formation time, but with reduced cross sections, proportional to
the number of original
valence quarks.  
The Pauli principle is applied
to BB collisions by blocking the final state 
if the outgoing phase space is occupied.
No Bose effects for mesons  are implemented in UrQMD. 

\section{UrQMD analysis of Au+Au collisions.}

Now let us use the UrQMD model to explore to which extent 
LTE can be reached at least in a small cell around the origin of 
central (b=0 fm)  Au+Au collisions
at 10.7 AGeV. 
The formation of hot and dense hadronic matter
is expected in this reaction,
therefore the comparison of local
system properties predicted by UrQMD
with those of the ideal hadron gas 
looks promising. 

For our analysis 
we consider a cubic cell of volume $V=5\times5\times 5$~fm$^3$
centered around the origin of center-of-mass of Au+Au
system. 
The collective velocity
of this cell is equal to zero because the system 
geometry  gives no preferable direction for 
the collective motion.
The cell size should be small enough to avoid
collective flow (and streaming) and not to underestimate 
densities inside the cell.
On the other hand it should be large enough
to contain enough particles inside the cell, which guarantees 
reasonably small fluctuations
of particle observables in the cell. 
From our analysis (see below) it appears that
a cubic cell with a side of 5~fm 
satisfies simultaneously both above requirements
for 200 Au+Au central collisions at AGS energy.

We start by  analyzing  the time evolution  
of different physical quantities in the Au+Au center-of-mass frame. 
Time $t=0$ fm/c corresponds to the moment when the
two Lorentz contracted nuclei touch each other.
The maximum overlap of the nuclei
occurs at about 2.5~fm/c. 
At time 
$t=2R_{Au}/(\gamma_{cm}v_{cm})\cong$5.5 fm/c 
the freely streaming nuclei would have passed through each other. 

The reaction is dominated by 
strong momentum anisotropies for all particles
inside the cell. 
During the entire high energy density phase, $t<$10~fm/c, 
the widths of the velocity distributions 
in the longitudinal ($z-) $direction,
$\sigma_z$,
are much larger then the widths in the transverse ($x-$, $y-$) directions,
$\sigma_x$ and $\sigma_y$, 
for each hadron species. 
However, 
for $t\geq$10~fm/c  
the anisotropies inside the central cell disappear and 
the widths $\sigma_i$  
become approximately equal.  The velocity component distributions
of nucleons in the cell, shown in 
Fig.~1 at two different times,
confirm this result. The shapes of the 
velocity 
distributions are  very different for $v_z$ and $v_{\perp}$  
at $t$=5~fm/c. They 
become very similar at $t$=10~fm/c.
This approximate isotropy 
of hadronic velocity distributions in the central cell appears to be 
due to the escape of fast or non-scattered particles from the cell 
and 
as a result of many hadronic rescatterings. 
At $t=$10~fm/c 
the newly produced particles in the cell  have undergone  
from 1.5 (pions and $\bar K$) to 5 (K) elastic collisions, while  
the baryons have suffered more than 20 strong interactions.
 Fig.~2 shows the large difference between the hadrons containing original
constituent quarks ($<n_{coll}>\cong 10-30$) and newly bred mesons 
($<n_{coll}>\cong 1-3$). 
There are still some collisions in the cell at $t>$18 fm/c.
Fig.~3 shows that for $t\leq$18~fm/c 
only a small fraction of the particles in the cell
is frozen out before that time.  

The isotropy of the momentum distributions for all hadron species
in the cell is a necessary prerequisite for LTE. 
The next
observation is that hadron energy spectra in the central cell at $t\geq$10 fm/c 
are nicely fitted by Boltzmann distributions, $\exp(-E_i/T)$,
with rapidly falling $T-$values vs. time. At each moment, however, 
all particle species~$i$ exhibit  practically the same slope.
The "temperatures", $T$, agree to within $\Delta T\cong$10~MeV.

\section{Ideal Gas Statistical Model.}

To study chemical equilibrium in detail,  
let us now compare the hadron spectra and yields in the cell, as
predicted by UrQMD model,  with those of 
a Statistical Model of an ideal hadron gas  (SM) \cite{Ogloblin,Halm}. 
For this comparison of the UrQMD results to the SM model
the time interval from 10 to 18~fm/c is  chosen.
The largest values of the energy density, $\varepsilon \cong 21\varepsilon_0$,
and baryonic density, $\rho_B\cong 10\rho_0$, in the central reaction zone 
are reached earlier, at $t\cong$4-5~fm/c. At this time, however, the system
of particles in the cell is not yet thermalized  and  still 
exhibits  strong velocity anisotropy.
At $t\cong 4-5$~fm/c there is also a large fraction of 
non-formed and non-interacting particles 
in the cell, particularly mesons (see Fig.~3).
Hence, the physical interpretation of
this early stage of the reaction should - in the present approach -
not be carried forward in a SM model. 
For $t\geq$10~fm/c the fraction of non-formed particles
goes to zero rapidly.   On the other hand 
the total number of particles in the cell is small 
at $t>18$~fm/c and the system approaches its thermal freeze-out 
(no collisions occur) stage. 
In this paper we restrict our analysis of LTE to the time interval 
between $t$=10--18~fm/c. We  do not discuss the ``initial'' stage
with high energy density and the ``final'' kinetic stage
of particle freeze-out. 

For a thermalized ideal gas 
the momentum spectrum of particle species ``i'' in 
the rest frame of the cell can be presented as
\begin{equation}
\frac {dN_i}{d^3p}~=~\frac{g_i~V}{(2\pi\hbar)^3}~
\exp\left(\frac{\mu_i}{T}\right)~
\exp\left(- \frac{E_i}{T}\right)~\equiv~\frac{g_i~V}{(2\pi\hbar)^3}~f_i~,
\label{1}
\end{equation}
where $V$ is the volume of the cell,
$T$ is the temperature of the system in the cell, 
$g_i$, $m_i$, $E_i=(p^2+m_i^2)^{1/2}$ and $\mu_i$ are 
the degeneracy factor, the mass, the energy and 
the chemical potential
of the hadron  species ``i'', correspondingly.    
Here,  
Boltzmann
distribution functions, $f_i$,  are used 
for all hadrons.  
Quantum statistical effects
are not included in the present analysis.

With the term thermodynamical equilibrium we specifically 
mean   both kinetic and chemical equilibrium.
The chemical potential of particle {\it i} in LTE can be represented 
as
\begin{equation}
\mu_i ~=~ b_i\mu_B~+~s_i\mu_S 
\label{2} 
\end{equation} 
in terms of  the
baryonic-, $\mu_B$, and strange-, $\mu_S$, 
chemical potentials ($b_i=0,\pm 1$, $s_i=0,\mp 1,\mp 2,\mp 3$ 
are, respectively, the 
baryonic- and strange- charges of the particle  species ``i'').
The electric
chemical potential considered in \cite{el1,el2} is neglected.

\section{Results and discussion.}

Here, 200 Au+Au events, generated with UrQMD, are analyzed 
using the following procedure. 
The average energy
density, $\varepsilon$, baryonic density, $\rho_B$, and net
strangeness density, $\rho_S$, in the cell  
are calculated for 10$\leq t\leq $18~fm/c.
The contributions of both formed and not yet formed
particles are included. 
The values of $\varepsilon$,
$\rho_B$ and $\rho_S$ 
are inserted in the l.h.s. of the
following equations:
\begin{equation}
\varepsilon~=~\sum_i \varepsilon_i~,~~~
\rho_B~=~\sum _i b_i~n_i~,~~~
\rho_S~=~\sum _i s_i~n_i~.
\label{3}
\end{equation}
Here
$$n_i~=~\frac{g_i}{2\pi ^2\hbar^3}~\int _0^{\infty}p^2dp~
f_i~,~~~ \varepsilon_i~=~\frac{g_i}{2\pi ^2\hbar^3}~\int _0^{\infty}p^2dp~
(p^2~+~m_i^2)^{1/2}~f_i~
$$
are the number- and energy density of particle species ``i'' predicted
by the ideal gas model.
The ideal gas model employs 
the same hadron species ``i'' as the UrQMD model. 
Solving the system of three equations (3) 
for the three unknown parameters,
$T$, $\mu_B$ and $\mu_S$, we find  the results  
presented in Table~1.

\subsection{Thermalization in central reaction zone.}

Now we can turn to the predictions of the Statistical model (SM)
for the specific single particle spectra, Eq.~(1),
by using the intensive parameters $T$, $\mu_B$ and $\mu_S$,
extracted from the average energy density, $\varepsilon$, baryon density, 
$\rho_B$,
and strangeness density, $\rho_S$. Also the pressure and the entropy 
can be calculated swiftly (see below).
The Boltzmann energy spectra Eq.~(1) predicted by the SM 
with $T,\mu_B,\mu_S$ values from  Table~1 and 
the spectra 
obtained in the UrQMD model in the central cell for
different particle species  agree surprisingly well 
for all instants $t$ between 10 and
18~fm/c. 
Fig.~4 shows  this comparison for $t$=13~fm/c. 
Between 10 and 18 fm/c also the hadronic yields, obtained in the SM,  
 are rather close to the UrQMD data, 
see 
Fig.~3. These results seem to indicate 
that the LTE is closely resembled for the hadronic matter 
in central cell of 
central Au+Au collisions at 10.7 AGeV 
at least for $t=$10 to  18 fm/c.
The statistical model parameters, i.e.  
the main thermodynamic characteristics of 
the cell change rapidly with time. This clearly demonstrates that 
a fireball type description of hadronic matter is inadequate. 

Table 1 shows that during this time interval (10 -- 18 fm/c) 
the baryonic density and temperature in the cell  change 
from $\rho_B\cong 2\rho_0$  to 0.3$\rho_0$  
($\rho_0 = 0.16$~fm$^{-3}$ is the normal
nuclear density) 
and from $T\cong$150~MeV
to 100~MeV. 
The time evolution of the central cell in ($T$, $\rho_B-$)  plane  
is shown
in 
Fig.~5. It resembles closely an isentropic hydrodynamical expansion.
Let us develop this idea in detail.

\subsection{Equation of State.}

The hadron pressure is given in the statistical model by 
\begin{equation}
P ~=~\sum_i \frac{g_i}{2\pi ^2\hbar^3}\int_0^{\infty}
p^2 dp~\frac{p^2}{3(p^2+m_i^2)^{1/2}}~f_i~.
\label{4}
\end{equation}
Using Eq.~(4) we find that the equation of state, 
$P(\varepsilon)$,  of
the hadronic matter in the central region of Au+Au collisions at 10.7 AGeV
is nearly linear with $\varepsilon$. In fact, $P(\varepsilon)/\varepsilon$ 
is constant for the whole time interval $t$=10--18~fm/c 
\begin{equation}
P/\varepsilon~\cong~0.12~.
\label{5}\end{equation}
This constant corresponds to the square of the speed of sound in the 
medium, $c_s^2=(dP/d\varepsilon)\cong$0.12, which is very similar
to that in a resonance gas, $c_s^2\cong$0.14,     
obtained  in \cite{Shuryak}. 
The constant behaviour of  $P(\varepsilon)/\varepsilon$  
confirms that transport models like UrQMD and RQMD do not
predict any phase transition which would be manifest in the softest
point of the EoS, $P(\varepsilon)/\varepsilon$   \cite{flow}.

In 
Fig.~6 we compare the ideal gas
pressure Eq.~(4) with longitudinal, 
$P^{mic}_{\{z\}}$, and transverse, $P^{mic}_{\{x,y\}}$,
pressures obtained in UrQMD as \cite{berenguer,sorge}:
\begin{equation}
P_{\{x,y,z\}}^{mic}~=\sum_h 
\frac{p^2_{h\{x,y,z\}}}{3V(m_h^2~+~p^2_h)^{1/2}}~,
\label{6}
\end{equation}
where V is volume of the cell, 
$p_h$ represents the particle momentum and the 
sum in Eq.~(6) is taken
over all hadrons $h$
in the cell. 
Fig.~6 shows a   strong
difference between $P_{\{z\}}^{mic}$ and $P_{\{x,y\}}^{mic}$
for $t<$10~fm/c. That this is  due to the longitudinal flow in the cell
at early times can be seen from the dashed curve, obtained for a cell with
a shortened 
longitudinal side, $\Delta z<1$ fm.  
The difference between $P_{\perp}$ and $P_{\parallel}$ vanishes already 
at $t=$6 fm/c, but at so early times there is still no equilibrium in the cell:
the slopes of the energy spectra are quite different for different 
particle species. Only at $t\cong$10  fm/c the hadronic matter 
in the cell becomes equilibrated. At this time 
the microscopic pressure Eq.(6) in both cells also 
becomes nearly isotropic and approximately equal
to the ideal gas one Eq.(4), 
$3P_{\{z\}}^{mic}\cong 3P_{\{x,y\}}^{mic}\cong P$.
The strong anisotropy of the transverse and the longitudinal components
of the pressure is also observed in three-fluid hydrodynamical model for the initial 
stages of central Au+Au collisions at AGS \cite{hydro}. This finding
is in stark 
contrast to the one-fluid
hydrodynamical model, where the pressure is assumed to be locally 
isotropic at all times.

Corse graining problems inhibit the direct microscopic calculations
of the entropy.
However, the entropy density $s$ can be calculated from the ideal gas model
(using the microscopic $\varepsilon$, $\rho_B$, $\rho_S$ as input,
 see above) 
by the thermodynamical identity 
$$ \varepsilon=Ts+\mu_B\rho_B+\mu_S\rho_S-P. $$
The strange particle contribution 
is negligibly small ($\mu_S\rho_S\approx1$~MeV maximally) 
in comparison to the other terms which are of order 100 MeV.
The entropy per baryon in the cell is nearly constant
and is equal to
\begin{equation}
S/A=s/\rho_B ~\cong~12~.
\label{7}\end{equation}
Hence, the time evolution of 
the cell is nearly isentropic 
for times $t$=10--18~fm/c  
and, therefore, 
similar to an ideal hydrodynamic expansion. 

\subsection{Comparison with infinite equilibrated matter.}

Quantum statistical effects in the ideal hadron gas
are small for the temperatures and chemical potentials, $T$ and $\mu$,
from the Table~1. Bose statistics 
gives the strongest effect  for the pions --
the Bose pion number density at $T$= 100--150~MeV is
about 10\% larger than the Boltzmann approximation.
The quantum statistical
effects are only 1-2\% for other 
hadrons. 

To make a conclusion about LTE in heavy ion collisions
in a given microscopic model one has to perform a comparison
not only to the Statistical Model, but also to an equilibrated
infinite matter system within the same microscopic model.
The energy spectra
and the particle  yields in the central cell of Au+Au collisions 
were compared to the equilibrated hadronic
matter calculations obtained in a   box 
with periodic boundary conditions by employing the UrQMD model
to such a closed system with the approximate microcanonical
infinite matter. This closed box calculations must 
clearly be distinguished from the open system, off equilibrium, 
in the central cell of heavy ion collisions discussed before. 
The 
particle yields and energy spectra predicted in the UrQMD  
box calculations \cite{box} for  each set of 
$\varepsilon$, $\rho_B$ and $\rho_S$
values taken from Table 1 for the different time steps 
are in a good agreement with
the UrQMD predictions in the central cell of Au+Au collisions 
(see Fig.~4). This additionally confirms that at the intermediate stage 
of the reaction $t=$10--18 fm/c the hadronic matter in the cell is in
thermodynamical equilibrium. 
This also means that  the closed box calculations 
(as well as SM) can be used to describe - of course not to 
predict - the time evolution of locally equilibrated open system of 
the heavy ion collisions discussed above.

The static box values agree approximately
with the ideal gas hadron model.
However, there are  enhancements
of the low-energy pion yields 
as compared to the ideal gas results
observed
in both the UrQMD cell- and the box-
calculations, seen in 
Fig.~4.
The box
calculations \cite{box}, at $\rho_B=\rho_0$, 
show that this enhancement
becomes even stronger with increasing  energy density.
This enhancement is due to the violation of the detailed balance
conditions for string- and many-body resonance decays   
in microscopic models like UrQMD \cite{box}. 
The additional pion production from many-body decay
channels acts like a
positive pion chemical potential. The Bose enhancement of pions 
should therefore be more    pronounced  at higher energy densities 
\cite{Mishustin}.

\section{Conclusions.}

Equilibration of hadronic
matter produced in the central cell of Au+Au collisions at 
10.7~AGeV has been studied 
in the microscopic
UrQMD model. It is found that 
\begin{itemize}
\item 
There are large deviations from thermal equilibrium for the high density phase, 
$t<$10~fm/c. 
\item 
Local thermal and chemical equilibration seems to set  in, in the central cell, for
 the expansion stage, $t$=10--18~fm/c. 
\item
Both the different particle yields and the energy spectra 
in the central cell change drastically with time. Still, their instantaneous
values agree, separately, for different time steps $t\geq$10~fm/c, 
with correspondingly adjusted ideal hadron gas model values. 
\item
The temperatures and baryonic densities extracted in the central 
cell drop rapidly from $T=$150 to 100~MeV and from $\rho_B=2\rho_0$ to 
$0.3\rho_0$ for $t=$10 to 18 fm/c. 
\item
The entropy per baryon
is nearly constant, $S/A=s/ \rho_B \cong 12$, during the time interval 
$t=$10--18 fm/c. 
\item
The equation of state of this complex mixture of different
particle "fluids" has a very simple form 
$P(\varepsilon)\cong 0.12\varepsilon$ for this range of 
$T,\rho_B,\rho_S-$values.
\item
A substantial enhancement  of low-energy pions  is observed 
in the UrQMD calculations as compared to the ideal gas results. This effect
increases at higher energy densities due to the string- and many-body decays
of the resonances.

\item
At $t$=10--18~fm/c the spectra and hadron multiplicities 
in the central reaction cell 
are in agreement with those found
for equilibrated infinite matter in UrQMD calculations in a 
box with periodic boundary conditions. 
The time scale of the thermal and chemical equilibration 
is, however, 
much shorter in the present calculations for nuclear reactions in  the cell  
than in the UrQMD box calculations. 
It is the short-lived pre-equilibrium high density initial phase which prepares
the extreme entropy densities forming doorway state to the quasi-isentropic 
expansion for $t\geq$10~fm/c.

\end{itemize}

Let us stress again that these  studies support a picture of  strong
deviation from {\it local}
thermodynamical equilibration in the central cell for $t<$10~fm/c in
A+A collisions at AGS energies. Large collective streaming-through
flow is found even in the central cell at early times.
The whole system never reaches {\it global}
thermodynamical equilibrium. However, local equilibration is
approached in central cell for the later times, $t\geq$10 fm/c.  
These flow effects are 
not large, however, at $t\geq$10  fm/c in the central cell,
that is why it was chosen. The
averaged squared velocities in the central cell at $t$=10~fm/c are 
$ \langle(v_z^{flow})^2\rangle \cong 
0.04$ and $ \langle (v_{x,y}^{flow})^2\rangle \cong 0.01$
for all particle species. These values
are much smaller
than the thermal velocities of the various 
hadrons in the cell at that time. For nucleons
one finds $\langle v_{th}^2 \rangle \cong 3T/m_N\cong 0.5$.
Thermal velocities for pions and kaons are even larger. 
Our analysis neglects this  small collective motion inside the central
cell at $t$=10--18~fm.
More careful studies give corrections
to this picture: in 
Fig.~4 one observes that the ``best fits''
to the particle spectra would give $T_{N,\Delta}>T_{\pi, K}$: small collective
motion inside the cell can imitate slightly larger
(several percents) effective temperatures
for heavier particles in the ideal gas calculations.  
Much larger flow velocities are found for all cells other than the central one.

A similar analysis has been done at lower energies in \cite{berenguer,Goren,Konopka}.
The next step is to study the approach to 
LTE for the earlier phase of the evolution
and at higher collision energies (from the AGS to the SPS).
This requires much smaller cell sizes and correspondingly
larger samples of A+A events. 

\section{Acknowledgements.}
L.V.B. is grateful to L. Csernai, I.N. Mishustin, E. Shuryak, 
J. Vaagen and E.E. Zabrodin for fruitful discussions.
M.I.G. is thankful to K.A. Bugaev and M. Ga\'zdzicki for useful comments.
This work was supported by A. v. Humboldt Stiftung,   
DFG, BMBF, Graduiertenkolleg "Exp. u. Theoret. Schwerionenphysik" and 
J. Buchmann Stiftung.

\newpage


\begin{table}
\caption{
The thermodynamic characteristics of the hadronic matter
in the central cell (V=125 fm$^3$) in central Au+Au collisions at 10.7AGeV.
The energy density, $\varepsilon$,  baryonic density,  $\rho_B$, 
and net strangeness density, $\rho_S$, are obtained from the 
microscopic UrQMD calculations. 
The temperature, $T$, baryonic chemical potential, $\mu_B$, and strange 
chemical potential, $\mu_S$, are extracted from the ideal hadron gas model,
using $\varepsilon$, $\rho_B$ and $\rho_S$ as input.
}

\vspace{0.5cm}
\begin{tabular}{ccccccc} 
\hline \hline
$t$ & $\varepsilon$ & $\rho_B$ & $\rho_S$ & $T$ & $\mu_B$ & $\mu_S$\\ 
(fm) & (MeV/fm$^3$)&(fm$^{-3}$)&(fm$^{-3}$)& (MeV)&(MeV)&(MeV)\\ 
\hline \hline
10 & 675 & 0.33 & -0.008 & 147 & 510 & 128  \\
11 & 511 & 0.26 & -0.007 & 140 & 519 & 122 \\
12 & 384 & 0.20 & -0.006 & 134 & 527 & 116 \\
13 & 293 & 0.16 & -0.004 & 128 & 534 & 112 \\
14 & 222 & 0.12 & -0.003 & 122 & 542 & 107 \\
15 & 170 & 0.10 & -0.003 & 116 & 553 & 101 \\
16 & 134 & 0.08 & -0.003 & 111 & 560 & 96 \\
17 & 102 & 0.06 & -0.003 & 106 & 567 & 87 \\
18 & 80 & 0.05 & -0.002 & 101 & 574 & 79 \\
\hline \hline
\end{tabular}
\label{tab1}
\end{table}

\newpage


\begin{figure}[htp]
\centerline{\epsfysize=13cm \epsfbox{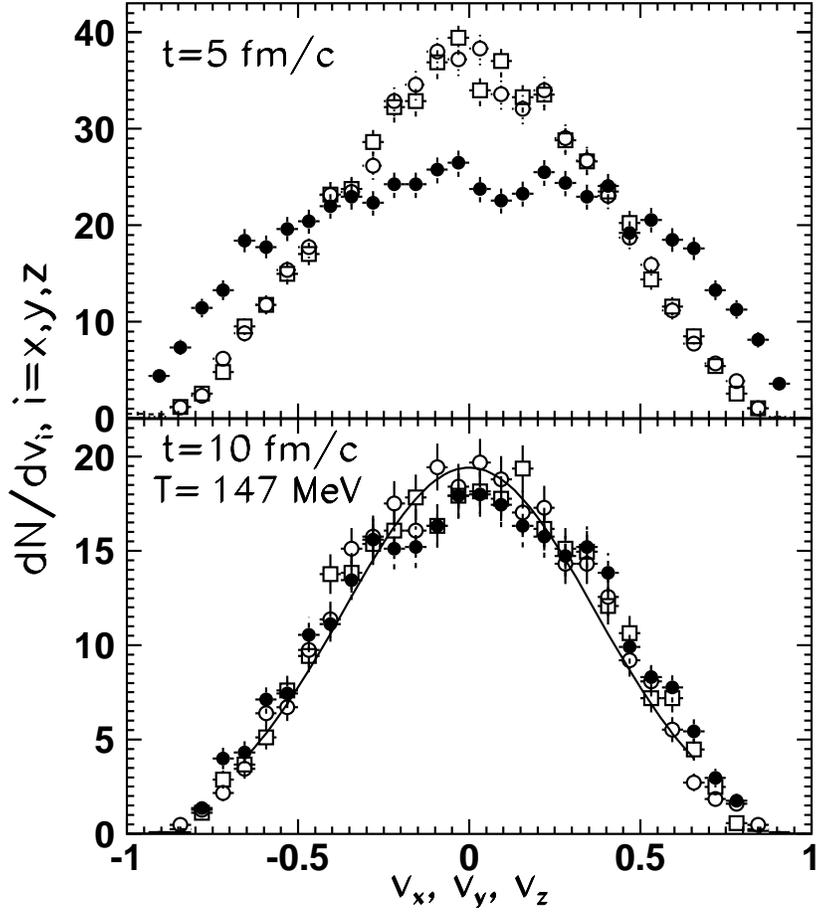}}
\caption{ 
Nucleon velocity distributions $dN/dv_i$ 
($i=z$ ($\bullet $), $x$ ($\Box$) and $y$ ($\bigcirc$)) 
in central cell of Au+Au collisions at 10.7 AGeV 
at $t=$5 (upper frame) and  10 (lower frame) fm/c. 
The solid line corresponds 
to the Maxwell-Boltzmann velocity distribution
$dN/dv_i \sim \exp(-m_Nv_i^2/2T)$
for the ``non-relativistic'' ($v_i<0.6$) nucleons  with temperature
$T=147$~MeV as 
obtained in ideal gas model. 
}
\label{fig1}
\end{figure}

\newpage

\begin{figure}[htp]
\centerline{\epsfysize=18cm \epsfbox{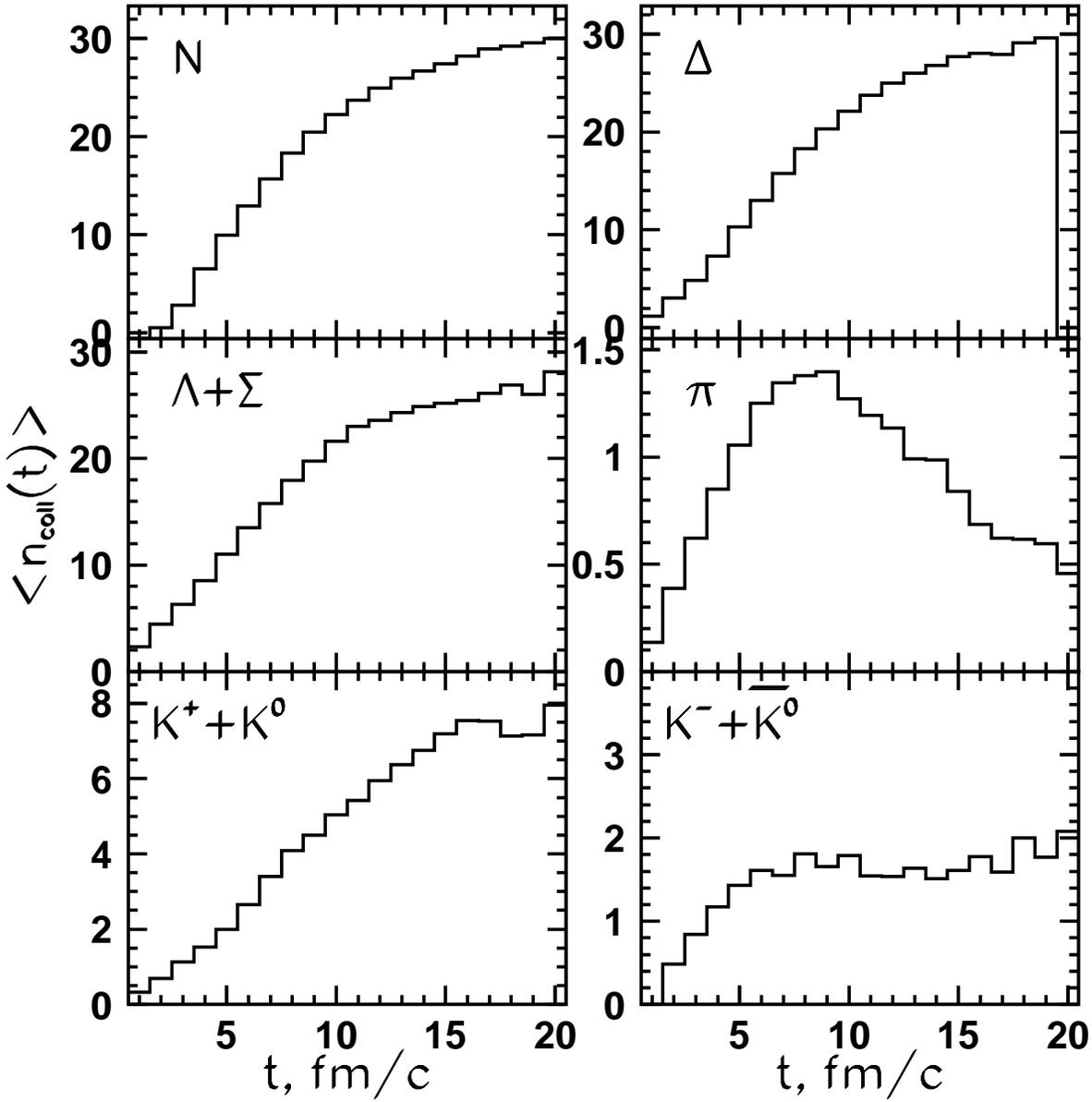}}
\caption{ 
The number of collisions, $<n_{coll}(t)>$, in the central cell 
of Au+Au collisions at 10.7 AGeV versus time, $t$, 
for nucleons, deltas, lambdas plus sigmas, pions, kaons and antikaons.
}
\label{fig2}
\end{figure}

\begin{figure}[htp]
\centerline{\epsfysize=18cm \epsfbox{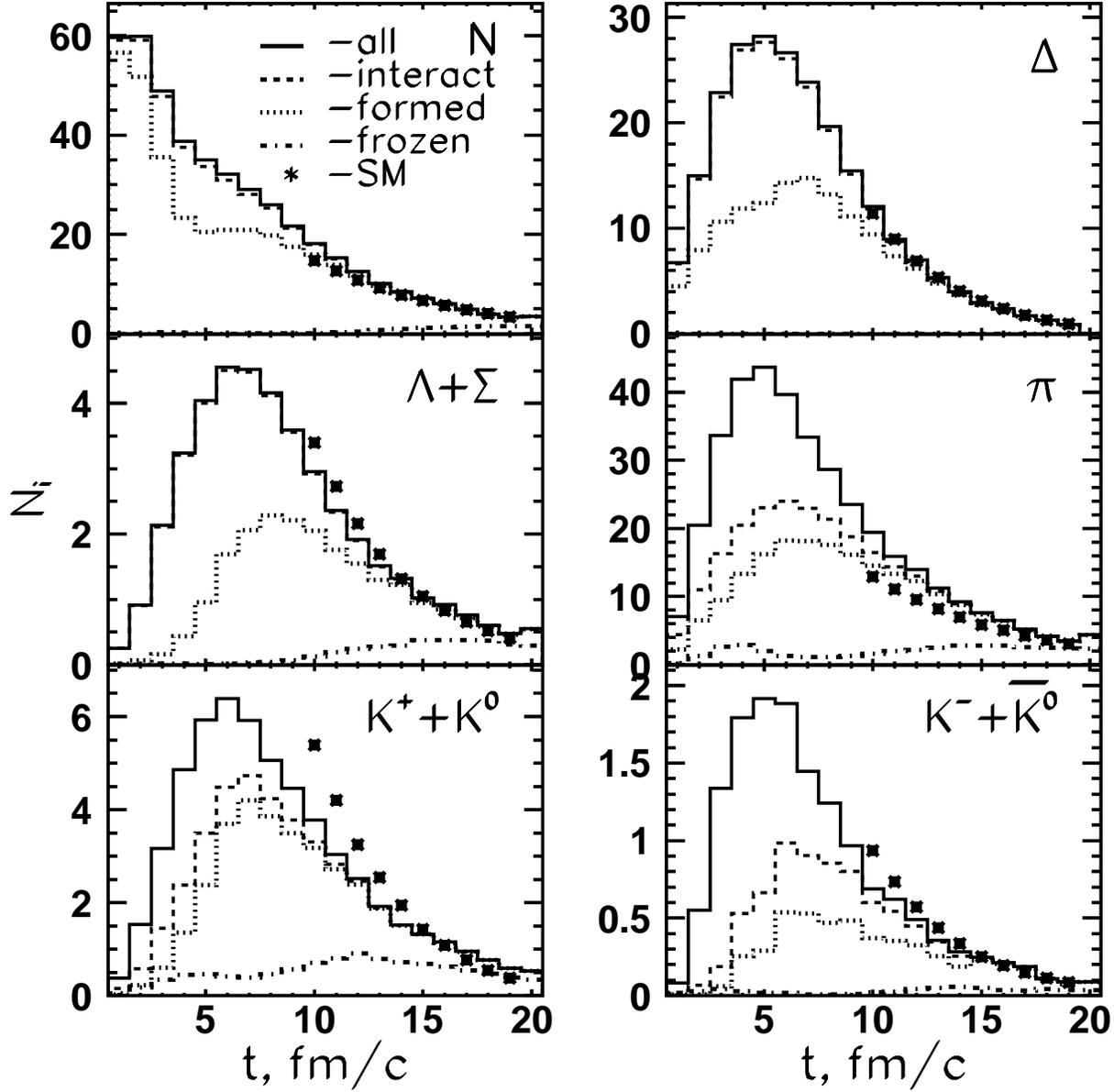}}
\caption{ 
The number of particles in the central cell of Au+Au collisions at 10.7 AGeV
as a function of time as obtained 
in  UrQMD model (histograms).
Solid lines correspond to all hadrons in the cell,
dashed lines -- 
to interacting particles and dotted lines -- 
to formed hadrons. The numbers of frozen particles in the cell  
are shown by dot-dashed lines. The points represent the predictions
of the ideal gas model as calculated at the same energy, baryonic and
strangeness densities as for the UrQMD central cell. 
}
\label{fig3}
\end{figure}

\begin{figure}[htp]
\centerline{\epsfysize=13cm \epsfbox{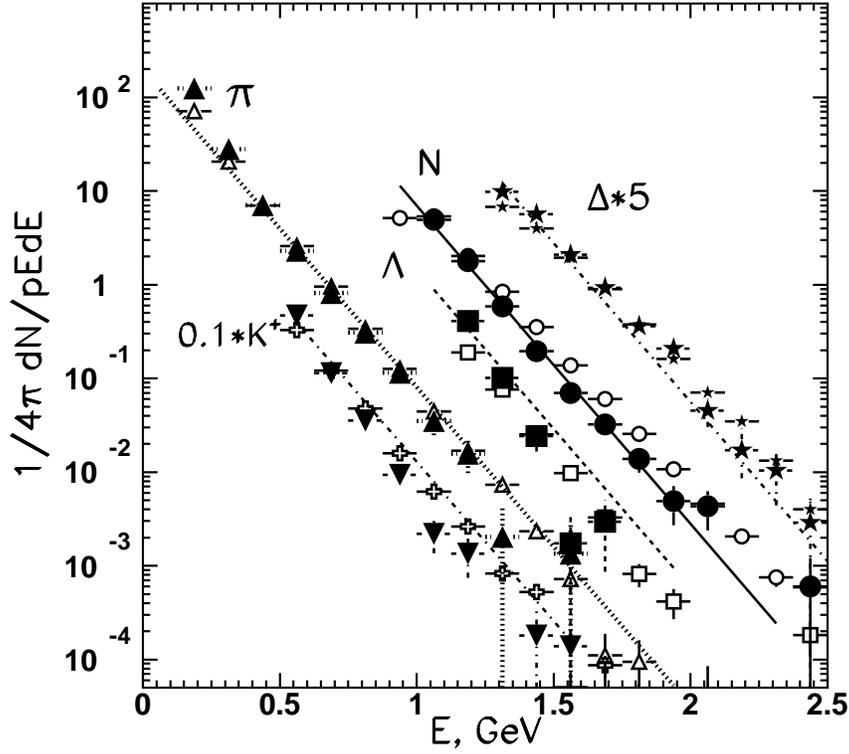}}
\caption{ 
Energy spectra of $N$ ($\bullet $), $\Lambda $ ($\Box $),
$\pi $ ($\bigtriangleup $), $K^+$ ($\bigtriangledown $) and $\Delta$ ($\star$)
in the central 125 fm$^3$ cell of Au+Au collisions at 10.7~A$\cdot$GeV at 
$t$=13~fm/c (black points) and in box calculations (open points) are fitted
by Boltzmann distributions, Eq.~(1) (lines)
using the parameters $T$=128~MeV, $\mu _B$=534~MeV, $\mu _S=112$~MeV, as   
obtained in the ideal gas model.
}
\label{fig4}
\end{figure}

\begin{figure}[htp]
\centerline{\epsfysize=18cm \epsfbox{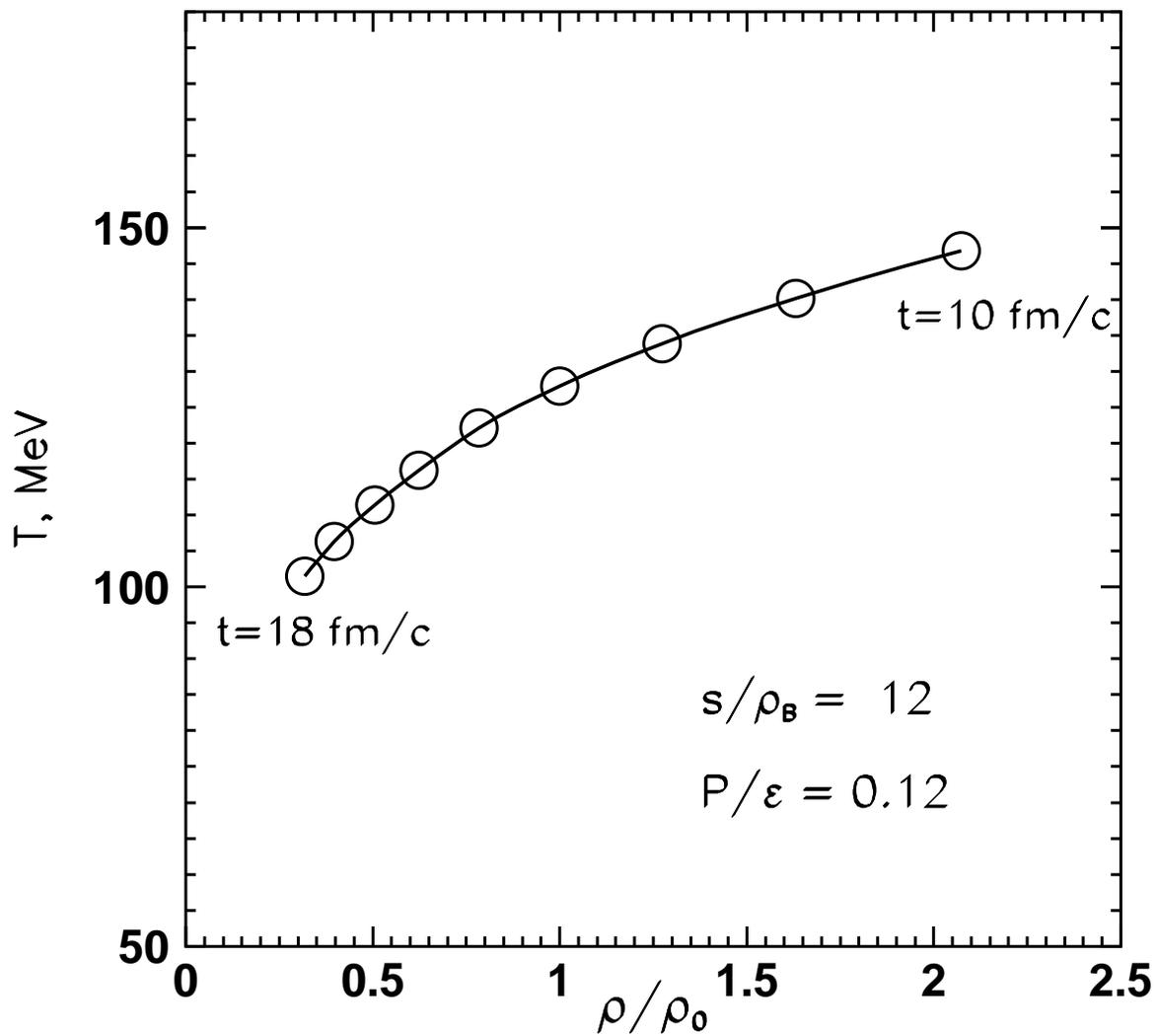}}
\caption{%
The evolution of the baryon density, $\rho_B$, and temperature, $T$,
in the central cell of Au+Au collisions at 10.7 AGeV at times
$t$=10--18 fm/c.
The entropy per baryon is constant $s/\rho_B \approx 12$
thus exhibiting quasi-isentropic expansion in this central cell.
The equation of state has the form
$P\approx 0.12\varepsilon$.
}
\label{fig5}
\end{figure}

\begin{figure}[htp]
\centerline{\epsfysize=17cm \epsfbox{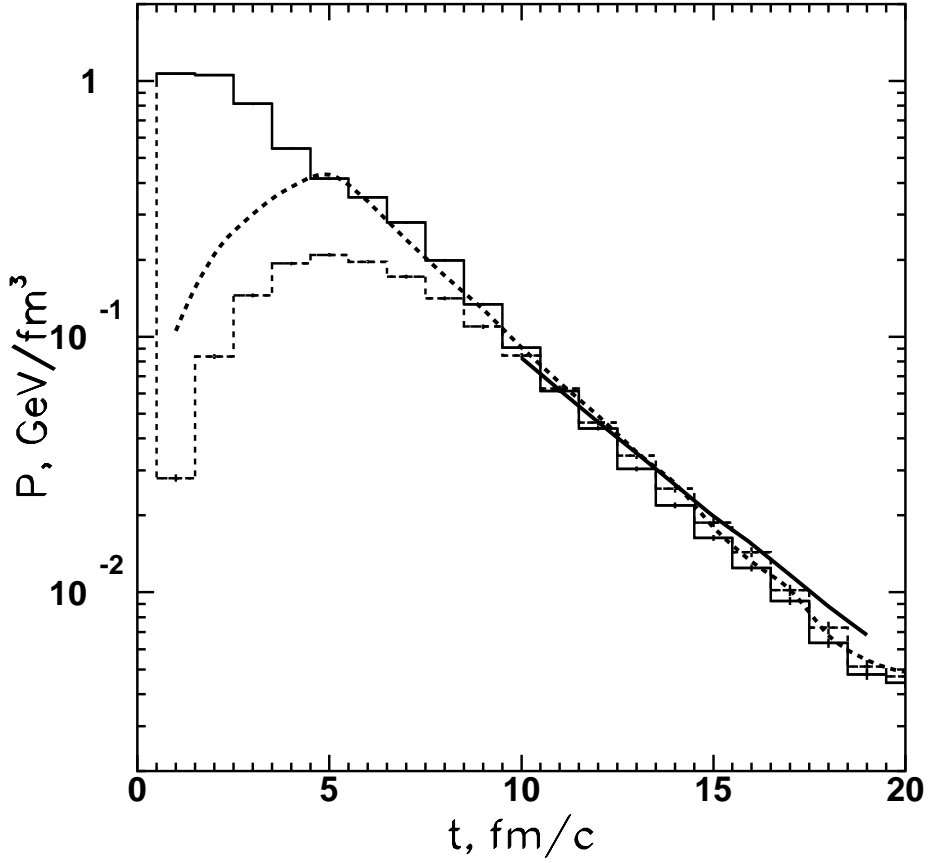}}
\caption{ 
The longitudinal ($3\cdot P_{\{z\}}$, solid histogram)
and the transverse ($3\cdot P_{\{x,y\}}$, dashed curves) 
diagonal 
components of the microscopic pressure  tensor 
in the central cell 
($\Delta x\!\times \!\Delta y\!\times \!\Delta z\!=$~
$\!5\!\times \!5\!\times \!5$~fm$^3$  (histograms) 
and $4\!\times \!4\!\times \!1$~fm$^3$, (dashed  line)) 
of Au+Au collisions at 10.7 AGeV calculated from 
the virial theorem Eq.~(6) are compared to 
the ideal gas pressure Eq.~(4) (solid line).
}
\label{fig6}
\end{figure}
\end{document}